# Optimal step-size of least mean absolute fourth algorithm in low SNR

Sihai Guan[1], Chun Meng[1], Bharat Biswal[1,2]

**Abstract:** There is a need to improve the capability of the adaptive filtering algorithm against Gaussian or multiple types of non-Gaussian noises, time-varying system, and systems with low SNR. In this paper, we propose an optimized least mean absolute fourth (OPLMF) algorithm, especially for a time-varying unknown system with low signal-noise-rate (SNR). The optimal step-size of OPLMF is obtained by minimizing the mean-square deviation (MSD) at any given moment in time. In addition, the mean convergence and steady-state error of OPLMF are derived. Also the theoretical computational complexity of OPLMF is analyzed. Furthermore, the simulation experiment results of system identification are used to illustrate the principle and efficiency of the OPLMF algorithm. The performance of the algorithm is analyzed mathematically and validated experimentally. Simulation results demonstrate that the proposed OPLMF is superior to the normalized LMF (NLMF) and variable step-size of LMF using quotient form (VSSLMFQ) algorithms.

**Keywords:** LMF; optimal step-size; non-Gaussian noises; time-varying; low SNR.

## 1 Introduction

Adaptive filter (AF) algorithms are frequently employed in linear systems [1], nonlinear systems [2], and distributed network systems [3], and has been used in many fields including biomedical engineering [4][5]. Recently, research has focused on AF algorithms based on high order error power (HOEP) conditions [6][7]. The most common HOEP algorithm are the least mean absolute third (LMAT) [7] and least mean fourth (LMF) algorithms, especially the LMF algorithm [6][10]-[14]. It is because the LMF algorithm outperforms the well-known least mean square (LMS) algorithm and achieves a better trade-off between the transient and steady-state performances of the adaptive filter [1][6]. More importantly, the LMF algorithm can achieve robust performance against the unknown noise of several different probability densities [1][6]. For example, impulsive noise has been modeled by an α-stable random process in previous studies [15][16]. It has been observed that the LMF yields better performance than that of the LMS algorithm for noise of sub-Gaussian nature, or noise with long-tailed probability density function [6][10]-[14]. However, the LMF algorithm has

Corresponding author: ✉Sihai Guan (gcihey@sina.com); ✉Chun Meng (chunmeng@uestc.edu.cn); ✉Bharat Biswal (bbiswal@gmail.com);

[1] School of Life science and technology, University of Electronic Science and Technology of China, No. 2006, Xiyuan Ave, West Hi-Tech Zone, Chengdu, Sichuan, 611731, China.

[2] Department of Biomedical Engineering, New Jersey Institute of Technology (NJIT), Newark, NJ, USA.

several stability problems that may limit its applications. For example, prior methods lead to suboptimal solutions in low signal-to-noise ratio (SNR) region, and are sensitive to the scaling of input signal and strong noise. Wallach and colleagues [6] showed that the stability of the algorithm about the Wiener solution depends upon both the adaptive filter input power and the noise power. To improve performance, normalized versions of the LMF algorithm have been proposed and studied [17]-[19]. None of these normalized LMF (NLMF) algorithms provides a global remedy to the above mentioned stability problems. However, algorithm stability depends upon the input signal power [19].

The step-size is a critical parameter in addressing the issue of obtaining either a fast convergence rate or low excess mean-square error [7]-[9]. A large step-size responds quickly to convergence, but may lead to a large mean-square deviation (MSD) and even cause loss of convergence, while a small step-size may degrade the tracking speed of the AF algorithm [7]. Therefore, the step-size must be chosen very carefully. Selection of the step-size should balance low steady-state error with a fast convergence rate. Thus, the variable step-size method attracts considerable attention in the field. There are various methods that can be adopted to get the step-size formula, such as the one previously developed by us that made use of a special function to update step-size [20]. In previous study we [7] utilized the various nonparametric variance estimates and proposed the nonparametric variable step-size LMAT algorithms. The approximating optimal step-size was selected in [8] such that the MSD was minimized. In the following study we [9] made use of MSD and proposed an optimized LMAT algorithm that addressed the optimization problem with the step-size. Similar method can be useful for the normalized LMF algorithm with the step-size. For LMF-type algorithms, there are several variable step-size LMF algorithms, such as NCLMF [21]. In this algorithm normalization does not protect the algorithm from divergence when the input power of the adaptive filter increases [22]. Therefore, motivated by Zhao and colleagues [23], Asad and colleagues [24] proposed a new variable step-size least-mean fourth algorithm by utilizing the quotient form (VSSLMFQ) in the step-size design, with the aim to achieve both robust and stable design. But this method still cannot tell us what optimal step-size of LMF. So, from the above discription, all the existing solutions are found to be either very sensitive to input and/or noise statistics or suffer from instability.

The effectiveness of the LMF-type algorithm are determined or affected by a number of factors including the feature of unknown time-varying system, the characteristics of the additive noise, low SNR and the input signal excitation. However, in practice, the measurement noise of an unknown system is non-Gaussian, and the system is often time-varying with low SNR. Therefore, on the basis of prior work presented in [9][10]-[14], an optimal step-size of the LMF algorithm (OPLMF) in this paper is derived such that the MSD is minimum at any given moment in time. The purpose of the OPLMF algorithm is to be able to deal with the different noise distributions including Gaussian, Uniform, Binary, Rayleigh and Poisson effectively. The mean convergence and MSD of the OPLMF algorithm are also derived in order to show that this algorithm is superior, theoretically. The computational complexity of the OPLMF algorithm is analyzed theoretically. Finally, we carried out seven system identification experiments to illustrate significant superiorities of the OPLMF algorithm over the NLMF and VSSLMFQ algorithms. Briefly, this paper is organized as follows. The proposed OPLMF algorithm is introduced in Section 2. Performance of the OPLMF algorithm is studied in Section 3. The numerical simulations are carried out in Section 4 and conclusions are presented in Section 5.

The main contributions of this work are as follows: (a) an OPLMF algorithm based on minimizing the MSD at any given moment in time, regardless the prob-density of noise. (b) Step-size of the OPLMF algorithm can achieve minimum steady-state error. (c) Stability of the proposed algorithm is studied. (d) Steady-state errors are derived from both Gaussian and non-Gaussian noises.

## 2 Proposed OPLMF algorithm

The coefficient vector of the unknown system is defined as $\mathbf{W_O}(n) = [w_o, w_1, w_2, \cdots, w_L]^T$. $L$ is the filter length. $\mathbf{X}(n) = [x(n), x(n+1), x(n+2), \cdots, x(n+L-1)]^T$ denotes the input data vector of the unknown system at times instant $n$, and $d(n)$ denotes the observed output signals, respectively.

$$d(n) = \mathbf{W_O}^T(n)\mathbf{X}(n) + \rho(n) \qquad (1)$$

where $\rho(n)$ is a stationary additive noise with zero mean and variance of $\sigma_\rho^2$. In addition, $\{\rho(n)\}$ is a stationary sequence of independent zero-mean random variables with a finite variance $\sigma_\rho^2$ and zero odd order moments and is assumed to be uncorrelated with any other signal. $\mathbf{X}(n)$ is also stationary with zero-mean, a variance of $\sigma_x^2$, and $\mathbf{X}(n)$ is Gaussian with a definite positive autocorrelation matrix $\mathbf{R} = E[\mathbf{X}(n)\mathbf{X}^T(n)]$.

The cost function used for obtaining the OPLMF algorithm is given by

$$\mathbf{W}(n) = \operatorname{argmin} J(\mathbf{W}(n)) = \operatorname{argmin} \frac{1}{4} E[e(n)^4] \qquad (2)$$

where

$$e(n) = d(n) - y(n) \qquad (3)$$

The corresponding filter output is

$$y(n) = \mathbf{W}^T(n)\mathbf{X}(n) \qquad (4)$$

Assuming

$$\mathbf{V}(n) = \mathbf{W_O}(n) - \mathbf{W}(n) \qquad (5)$$

So

$$e(n) = \mathbf{W_O}^T(n)\mathbf{X}(n) + \rho(n) - \mathbf{W}^T(n)\mathbf{X}(n) = \mathbf{V}^T(n)\mathbf{X}(n) + \rho(n) \qquad (6)$$

The updated recursion of coefficients vectors can be derived as Eq. (7):

$$\mathbf{W}(n+1) = \mathbf{W}(n) - \frac{\partial J(\mathbf{W}(n))}{\partial \mathbf{W}(n)} = \mathbf{W}(n) + \mu \mathbf{X}(n) e(n)^3 \qquad (7)$$

In Eq. (7), $\mu$ denotes the step-size [9]. The purpose of this work is to find an optimal step-size of variable LMF algorithm.

Combine Eq. (5) and Eq. (7),

$$\mathbf{V}(n+1) = \mathbf{V}(n) + \mu \mathbf{X}(n) e(n)^3 \qquad (8)$$

Based on [19], the following sufficient condition of convergence of $E[\mathbf{V}(n)]$ of the

LMF algorithm is derived as Eq. (9).

$$0 < \mu < \frac{1}{2E[\rho^2(n)]E[\|\mathbf{X}(n)\|^2]} \tag{9}$$

Eq. (9) implies that the stability of the LMF algorithm depends on the input power of the adaptive filter and the noise power. However, the Eq. (9) is derived under the assumption that the deviation vector is close to zero, i.e.,

$$\lim_{n \to \infty} \mathbf{V}(n+1) = \mathbf{0} \tag{10}$$

Premultiplying Eq. (8) by its transpose, using Eq. (5), taking the expected value and using the statistical properties of $\rho(n)$ leads to

$$\begin{aligned}
&E[\mathbf{V}^T(n+1)\mathbf{V}(n+1)] \\
&= E[(\mathbf{V}(n) + \mu\mathbf{X}(n)e(n)^3)^T(\mathbf{V}(n) + \mu\mathbf{X}(n)e(n)^3)] \\
&= \mathbf{E}[(\mathbf{V}^T(n) + \mu\mathbf{X}^T(n)e(n)^3)(\mathbf{V}(n) + \mu\mathbf{X}(n)e(n)^3)] \\
&= \mathbf{E}[\mathbf{V}^T(n)\mathbf{V}(n)] - 2\mu\mathbf{E}\{[\mathbf{X}^T(n)\mathbf{V}(n)]^4\} - 6\mu\mathbf{E}\{[\rho(n)]^2\}\mathbf{E}\{[\mathbf{X}^T(n)\mathbf{V}(n)]^2\} \\
&\quad + \mu^2\mathbf{E}\{[\mathbf{X}^T(n)\mathbf{V}(n)]^6\mathbf{X}^T(n)\mathbf{X}(n)\} + 15\mu^2\mathbf{E}\{[\rho(n)]^2\}\mathbf{E}\{[\mathbf{X}^T(n)\mathbf{V}(n)]^4\mathbf{X}^T(n)\mathbf{X}(n)\} \\
&\quad + 15\mu^2\mathbf{E}\{[\rho(n)]^4\}\mathbf{E}\{[\mathbf{X}^T(n)\mathbf{V}(n)]^2\mathbf{X}^T(n)\mathbf{X}(n)\} + \mu^2\mathbf{E}\{[\rho(n)]^6\}\mathbf{E}\{\mathbf{X}^T(n)\mathbf{X}(n)\} \\
&= \mathbf{E}[\mathbf{V}^T(n)\mathbf{V}(n)] - 2\mu\mathbf{E}\{[\mathbf{X}^T(n)\mathbf{V}(n)]^4\} - 6\mu\mathbf{E}\{[\rho(n)]^4\}\mathbf{E}\{[\mathbf{X}^T(n)\mathbf{V}(n)]^2\} \\
&\quad + \mu^2\mathbf{E}\{[\mathbf{X}^T(n)\mathbf{V}(n)]^6\mathbf{X}^T(n)\mathbf{X}(n)\} + 15\mu^2\mathbf{E}\{[\rho(n)]^2\}\mathbf{E}\{[\mathbf{X}^T(n)\mathbf{V}(n)]^4\mathbf{X}^T(n)\mathbf{X}(n)\} \\
&\quad + 15\mu^2\mathbf{E}\{[\rho(n)]^4\}\mathbf{E}\{[\mathbf{X}^T(n)\mathbf{V}(n)]^2\mathbf{X}^T(n)\mathbf{X}(n)\} + \mu^2L\sigma_x^2\mathbf{E}\{[\rho(n)]^6\}
\end{aligned} \tag{11}$$

So, we will derive Eq. (11) by giving each term,

$$\begin{aligned}
2\mu\mathbf{E}\{[\mathbf{X}^T(n)\mathbf{V}(n)]^4\} &\approx 6\mu\sigma_x^4\mathbf{E}[\mathbf{V}^T(n)\mathbf{V}(n)\mathbf{V}^T(n)\mathbf{V}(n)] \\
&= 6\mu\sigma_x^4\{\mathbf{E}[\mathbf{V}^T(n)\mathbf{V}(n)]\}^2
\end{aligned} \tag{12}$$

and

$$\begin{aligned}
6\mu\mathbf{E}\{[\rho(n)]^4\}\mathbf{E}\{[\mathbf{X}^T(n)\mathbf{V}(n)]^2\} &= 6\mu\mathbf{E}\{[\rho(n)]^4\}\mathbf{E}\{E[\mathbf{X}^T(n)\mathbf{V}(n)]^2|\mathbf{V}(n)]\} \\
&= 6\mu\mathbf{E}\{[\rho(n)]^4\}\mathbf{E}\{E[\mathbf{V}^T(n)\mathbf{X}(n)\mathbf{X}^T(n)\mathbf{V}(n)|\mathbf{V}(n)]\} \\
&= 6\mu\mathbf{E}\{[\rho(n)]^4\}\mathbf{E}\{\mathbf{V}^T(n)\mathbf{E}[\mathbf{X}(n)\mathbf{X}^T(n)|\mathbf{V}(n)]\mathbf{V}(n)\} \\
&\approx 6\mu\mathbf{E}\{[\rho(n)]^4\}\mathbf{E}\{\mathbf{V}^T(n)\mathbf{R}\mathbf{V}(n)\} \\
&= 6\mu\mathbf{E}\{[\rho(n)]^4\}\mathbf{E}\{\sigma_x^2\mathbf{V}^T(n)\mathbf{V}(n)\} \\
&\approx 6\mu\sigma_x^2\mathbf{E}\{[\rho(n)]^4\}\mathbf{E}[\mathbf{V}^T(n)\mathbf{V}(n)]
\end{aligned} \tag{13}$$

and

$$\begin{aligned}
\mu^2\mathbf{E}\{[\mathbf{X}^T(n)\mathbf{V}(n)]^6\mathbf{X}^T(n)\mathbf{X}(n)\} &= \mu^2\mathbf{E}\{E[\mathbf{X}^T(n)\mathbf{V}(n)]^6\mathbf{X}^T(n)\mathbf{X}(n)|\mathbf{V}(n)]\} \\
&= \mu^2 tr\{\mathbf{E}[\mathbf{X}^T(n)\mathbf{V}(n)]^6\mathbf{X}(n)\mathbf{X}^T(n)]\} \\
&= \mu^2\mathbf{E}\{tr(\mathbf{E}[\mathbf{X}^T(n)\mathbf{V}(n)]^6\mathbf{X}(n)\mathbf{X}^T(n)|\mathbf{V}(n)])\} \\
&\approx \mu^2\mathbf{E}\{(15L + 90)\sigma_x^8\mathbf{V}^T(n)\mathbf{V}(n)\mathbf{V}^T(n)\mathbf{V}(n)\mathbf{V}^T(n)\mathbf{V}(n)\} \\
&\approx \mu^2(15L + 90)\sigma_x^8\mathbf{E}^3[\mathbf{V}^T(n)\mathbf{V}(n)]
\end{aligned} \tag{14}$$

and

$$\begin{aligned}
&15\mu^2\mathbf{E}\{[\rho(n)]^2\}\mathbf{E}\{[\mathbf{X}^T(n)\mathbf{V}(n)]^4\mathbf{X}^T(n)\mathbf{X}(n)\} \\
&\approx 15\mu^2(3L + 12)\sigma_x^6\mathbf{E}\{[\rho(n)]^2\}\mathbf{E}^2[\mathbf{V}^T(n)\mathbf{V}(n)]
\end{aligned} \tag{15}$$

and

$$\begin{aligned}
&15\mu^2\mathbf{E}\{[\rho(n)]^4\}\mathbf{E}\{[\mathbf{X}^T(n)\mathbf{V}(n)]^2\mathbf{X}^T(n)\mathbf{X}(n)\} \\
&\approx 15\mu^2(L + 2)\sigma_x^4\mathbf{E}\{[\rho(n)]^4\}\mathbf{E}[\mathbf{V}^T(n)\mathbf{V}(n)]
\end{aligned} \tag{16}$$

Taking Eq. (11) ~ Eq. (16) into Eq. (11), Eq. (17) can be obtained.

$$E[\mathbf{V}^T(n+1)\mathbf{V}(n+1)]$$
$$= \mathbf{E}[\mathbf{V}^T(n)\mathbf{V}(n)] - 6\mu\sigma_x^4\{\mathbf{E}[\mathbf{V}^T(n)\mathbf{V}(n)]\}^2 - 6\mu\sigma_x^2\sigma_\rho^2\mathbf{E}[\mathbf{V}^T(n)\mathbf{V}(n)]$$
$$+(15L+90)\mu^2\sigma_x^8\mathbf{E}^3[\mathbf{V}^T(n)\mathbf{V}(n)] + 15(3L+12)\mu^2\sigma_x^6\sigma_\rho^2\mathbf{E}^2[\mathbf{V}^T(n)\mathbf{V}(n)]$$
$$+15(L+2)\mu^2\sigma_x^4\mathbf{E}\{[\rho(n)]^4\}\mathbf{E}[\mathbf{V}^T(n)\mathbf{V}(n)] + \mu^2 L\sigma_x^2\mathbf{E}\{[\rho(n)]^6\} \tag{17}$$

The MSD at times instant $n$ is defined as $\text{MSD}(n) = \text{E}[\mathbf{V}^T(n)\mathbf{V}(n)]$. So Eq. (17) can be derived as Eq. (18).

$$\text{MSD}(n+1) = \text{MSD}(n)\{1 + 15(L+2)\mu^2\sigma_x^4\mathbf{E}[\rho^4(n)] - 6\mu\sigma_x^2\sigma_\rho^2\}$$
$$+\text{MSD}(n)^2\{15(3L+12)\mu^2\sigma_x^6\sigma_\rho^2 - 6\mu\sigma_x^4\}$$
$$+\text{MSD}(n)^3(15L+90)\mu^2\sigma_x^8$$
$$+\mu^2 L\sigma_x^2\mathbf{E}[\rho^6(n)]$$
$$= f(L,\mu,\sigma_x,\rho(n))\text{MSD}(n) + g(L,\mu,\sigma_x)\text{MSD}(n)^2$$
$$+ (15L+90)\mu^2\sigma_x^8\text{MSD}(n)^3 + t(L,\mu,\sigma_x,\rho(n))$$
$$\tag{18}$$

where
$$f(L,\mu,\sigma_x,\rho(n)) = 1 + 15(L+2)\mu^2\sigma_x^4\mathbf{E}[\rho^4(n)] - 6\mu\sigma_x^2\sigma_\rho^2 \tag{20}$$

$$g(L,\mu,\sigma_x) = \{15(3L+12)\mu^2\sigma_x^6\sigma_\rho^2 - 6\mu\sigma_x^4\} \tag{19}$$

$$t(L,\mu,\sigma_x,\rho(n)) = \mu^2 L\sigma_x^2\mathbf{E}[\rho^6(n)] \tag{21}$$

When the algorithm tends to be stable, $\text{MSD}(n)$ is very small. So, Eq. (18) can be rewritten as Eq. (19) when neglect the higher order moments of $\text{MSD}(n)$

$$\text{MSD}(n+1) = f(L,\mu,\sigma_x,\rho(n))\text{MSD}(n) + t(L,\mu,\sigma_x,\rho(n)) \tag{22}$$

In the real practical world $\sigma_x^2$ are unknown [9], however, we can estimate $\sigma_x^2$ by using Eq. (23).
$$\sigma_x^2(n) = \gamma\sigma_x^2(n-1) + (1-\gamma)\mathbf{X}^T(n)\mathbf{X}(n) \tag{23}$$
where $\gamma$ is a small positive number to guarantee that the denominator of Eq. (23) remains finite when $\sigma_x^2(n) = 0$.

Based on NPVSS-NLMS algorithm [25], we can get $\left(1 - \frac{1}{2L}\right) \leq \gamma < 1$.

The result from Eq. (19) illustrates a "separation" between the convergence and misadjustment components. Therefore, the term $f(L,\mu,\sigma_x,\rho(n))$ influences the convergence rate of the algorithm. It can be noticed that the fastest convergence mode is obtained when the function of Eq. (19) reaches its minimum.

$$\mu(n) = \frac{\sigma_\rho^2}{5(L+2)\sigma_x^2 \mathbf{E}[\rho^4(n)]} \tag{24}$$

At this value, so to get $f(L,\mu,\sigma_x,\rho(n)) > 0$.

The stability condition can be found by imposing $|f(L,\mu,\sigma_x,\rho(n))| < 1$, which leads to Eq. (25).

$$0 < \mu(n) < \frac{\sigma_\rho^2}{5(L+2)\sigma_x^2 \mathbf{E}[\rho^4(n)]} \tag{25}$$

There are two important issues to be considered: (1) in the context of system identification, it is reasonable to follow a minimization problem in terms of the system misalignment; (2) we have main parameters $\mu$ which influences the overall performance of the OPLMF algorithm.

The derivative of $\mu$ is calculated at both ends of Eq. (18) at the same time.

$$\frac{\partial \mathrm{MSD}(n+1)}{\partial \mu} = \frac{\partial \mathrm{MSD}(n)}{\partial \mu} f(L,\mu,\sigma_x,\rho(n)) + \frac{\partial f(L,\mu,\sigma_x,\rho(n))}{\partial \mu}\mathrm{MSD}(n)$$

$$+ \frac{\partial \mathrm{MSD}(n)^2}{\partial \mu} g(L,\mu,\sigma_x) + \frac{\partial g(L,\mu,\sigma_x)}{\partial \mu}\mathrm{MSD}(n)^2$$

$$+ \frac{\partial \mathrm{MSD}(n)^3}{\partial \mu}(15L+90)\mu^2\sigma_x^8 + \frac{\partial\big((15L+90)\mu^2\sigma_x^8\big)}{\partial \mu}\mathrm{MSD}(n)^3$$

$$+ \frac{\partial t(L,\mu,\sigma_x,\rho(n))}{\partial \mu}$$

$$\tag{26}$$

Our goal is to meet Eq. (27).

$$\frac{\partial \mathrm{MSD}(n)}{\partial \mu} = 0 \tag{27}$$

Substituting Eq. (16) into Eq. (26), the optimal step-size is then given by

$$\frac{\partial f(L,\mu,\sigma_x,\rho(n))}{\partial \mu}\mathrm{MSD}(n) + \frac{\partial g(L,\mu,\sigma_x)}{\partial \mu} + \frac{\partial\big((15L+90)\mu^2\sigma_x^8\big)}{\partial \mu} + \frac{\partial t(L,\mu,\sigma_x,\rho(n))}{\partial \mu} = 0$$

$$\tag{28}$$

The optimal step-size is then given by

$$\mu(n) = \frac{3\mathrm{MSD}(n)\big(\sigma_\rho^2 + \sigma_x^2 \mathrm{MSD}(n)\big)}{\mu_f} \tag{29}$$

where $\mu_{f,n} = 15\sigma_x^2\mathrm{MSD}(n)\{(L+2)\mathbf{E}[\rho^4(n)] + (3L+12)\sigma_x^2\sigma_\rho^2\mathrm{MSD}(n) + (L+6)\sigma_x^4\mathrm{MSD}(n)^2\} + L\mathbf{E}[\rho^6(n)]$.

Using Eq. (29) in Eq. (18), followed by several computations, resulting in

$$\mathrm{MSD}(n+1) = \mathrm{MSD}(n)\left\{1 + \frac{15(9L+18)\mathrm{MSD}(n)^2\big(\sigma_\rho^2 + \sigma_x^2\mathrm{MSD}(n)\big)^2 \sigma_x^4 \mathbf{E}[\rho^4(n)]}{(\mu_{f,n})^2}\right.$$

$$\left. - \frac{18\mathrm{MSD}(n)\big(\sigma_\rho^2 + \sigma_x^2\mathrm{MSD}(n)\big)\sigma_x^2\sigma_\rho^2}{\mu_{f,n}}\right\} + \frac{9L\sigma_x^2\mathrm{MSD}(n)^2\big(\sigma_\rho^2 + \sigma_x^2\mathrm{MSD}(n)\big)^2 \mathbf{E}[\rho^6(n)]}{(\mu_{f,n})^2}$$

(30)

Denoting $\lim_{n\to\infty}\text{MSD}(n+1) = \lim_{n\to\infty}\text{MSD}(n) = \text{MSD}(\infty)$ and developing Eq. (30), we obtain the equation Eq. (31).

$$3\text{MSD}(\infty)^2\left\{\left(\frac{3L\sigma_x^2\mathbf{E}[\rho^6(n)]+5(9L+18)\sigma_x^4\mathbf{E}[\rho^4(n)]}{(\mu_{f,\infty})^2}\right)\left(\sigma_\rho^2+\sigma_x^2\text{MSD}(\infty)\right)\text{MSD}(\infty) - \frac{9\sigma_\rho^2}{\mu_{f,\infty}}\right\}\left(\sigma_\rho^2+\sigma_x^2\text{MSD}(\infty)\right) = 0 \quad (31)$$

Combine Eq. (23), we know $\text{MSD}(\infty) = 0$. It means that our aforementioned step-size adjusting mechanism can achieve zero steady-state mean-square deviation. For a Gaussian desired signal, $\mathbf{E}[\rho^4(n)] = 3\sigma_\rho^4$, $\mathbf{E}[\rho^6(n)] = 15\sigma_\rho^6$ [7][9][26]. For a Uniform desired signal, $\mathbf{E}[\rho^4(n)] = \frac{27}{7}\sigma_\rho^4$, $\mathbf{E}[\rho^6(n)] = \frac{9}{5}\sigma_\rho^6$ [7][9][26]. For a Rayleigh desired signal, $\mathbf{E}[\rho^4(n)] = 8\sigma_\rho^4$, $\mathbf{E}[\rho^6(n)] = 48\sigma_\rho^6$ [7][9][26]. For an Binary desired signal, $\mathbf{E}[\rho^4(n)] = \sigma_\rho^4$, $\mathbf{E}[\rho^6(n)] = \sigma_\rho^6$ [26]. For a Poisson desired signal (parameter $=\lambda$), $\mathbf{E}[\rho^4(n)] = \lambda + 7\lambda^2 + 6\lambda^3 + \lambda^4$, $\mathbf{E}[\rho^6(n)] = \lambda + 31\lambda^2 + 90\lambda^3 + 65\lambda^4 + 15$ [26].

The excess mean-square error (EMSE) at times instant $n$ is given by Eq. (32).

$$\begin{aligned}\text{EMSE}(n) &= \mathbf{E}[e^2(n)] \\ &= \mathbf{E}\{[\mathbf{V}^T(n)\mathbf{X}(n)+\rho(n)][\mathbf{X}^T(n)\mathbf{V}(n)+\rho(n)]\} \\ &= \sigma_\rho^2 + \sigma_x^2\text{MSD}(n) = \sigma_\rho^2\end{aligned} \quad (32)$$

Finally, the steady-state EMSE, $\text{EMSE}(\infty)$ is given by using Eq. (31) and Eq. (32).

$$\text{EMSE}(\infty) = \sigma_\rho^2 \quad (33)$$

Summary of the procedure for the OPLMAT algorithm based on the analysis presented above is given in Table 1.

Table1. OPLMF algorithm summary

| |
|---|
| Initialization: $\mathbf{W}(n) = \mathbf{0}, \gamma$ |
| Update: |
| $\sigma_x^2(n) = \gamma\sigma_x^2(n-1) + (1-\gamma)\mathbf{X}^T(n)\mathbf{X}(n)$ |
| $\mu(n) = \dfrac{3\text{MSD}(n)\left(\sigma_\rho^2+\sigma_x^2\text{MSD}(n)\right)}{15\sigma_x^2\text{MSD}(n)\{(L+2)\mathbf{E}[\rho^4(n)]+(3L+12)\sigma_x^2\sigma_\rho^2\text{MSD}(n)+(L+6)\sigma_x^4\text{MSD}(n)^2\}+L\mathbf{E}[\rho^6(n)]}$ |
| $\text{MSD}(n+1) = \text{MSD}(n)\{1+15(L+2)\mu^2(n)\sigma_x^4\mathbf{E}[\rho^4(n)] - 6\mu(n)\sigma_x^2\sigma_\rho^2\}$ $\qquad\qquad\qquad + \mu^2(n)L\sigma_x^2\mathbf{E}[\rho^6(n)]$ |
| $\mathbf{W}(n+1) = \mathbf{W}(n) - \dfrac{\partial J(\mathbf{W}(n))}{\partial \mathbf{W}(n)} = \mathbf{W}(n) + \mu(n)\mathbf{X}(n)e(n)^3$ |

## 3 Computational complexity

Table 2 shows the comparison of the computational complexity that is computed on the basis of the number of multiplications, divisions, comparisons, and pluses in each LMF-type algorithm's update equation, where $L$ is the tap length, and $N$ denotes the number of the input vector for the VSSLMFQ [24]. There is recursion required to compute $\mathbf{X}^T(n)\mathbf{X}(n)$ in OPLMF and NLMF. In addition, compared to NLMF, the updated step-size and $\sigma_x^2(n)$ formulas are added to the OPLMF algorithm, meaning that the computational complexity of OPLMF is greater than that of NLMF. However, the computational complexity of the OPLMF algorithm is less than the complexity of VSSLMFQ. From Table 2, the OPLMF algorithm has a considerable complexity advantage over VSSLMFQ.

Table 2. The computational complexity of OPLMF, NLMF and VSSLMFQ algorithms

| Algorithm | × | ÷ | Comparisons | + |
|---|---|---|---|---|
| NLMF [13] | $2L + 3$ | 0 | 0 | $3L$ |
| VSSLMFQ [24] | $2L + 11 + 2O(N)$ | 1 | 3 | $L + 3 + 2O(N)$ |
| OPLMF (proposed) | $2L + 16$ | 0 | 0 | $2L + 3$ |

where, "×" denotes Multiplications. "÷" denotes Divisions. "+" denotes Additions. "$N$" denotes length of input signal.

## 4 Simulation results

This section presents the results of simulations in the context of system identification using various noise distributions of both stationary and non-stationary systems to illustrate the accuracy and robustness of the OPLMF algorithm. The length of the unknown coefficient vector $\mathbf{W}_O(n)$ is assumed to be $L = 5$. The input signal $\mathbf{X}(n)$ is a Gaussian white noise with zero mean and $\sigma_x^2 = 1$. The correlated input signal $y(n)$ is calculated by using $y(n) = 0.5y(n) + x(n)$. In all of our experiments, the coefficient vectors are initialized as zero vectors. Five different noise distributions (Gaussian, Uniform, Binary, Rayleigh, and Passion) are used in the experiments. In order to verify the robustness of the pair of algorithms (for different low SNR, system time-varying characteristics, input signal type, and different types noise), we randomly select different low SNR, system characteristics, the input signal and noise type is combined into 7 groups experiment. $MSD(n) = 10log10(\|\mathbf{W}_O(n) - \mathbf{W}(n)\|_2^2)$ is used to measure the performance of algorithms. In addition, MSD error equals the absolute value of the difference between the simulation values and theoretical value (Eq. (26)). The results are obtained via Monte Carlo simulation using 50 independent run sets and an iteration number of 5000. The values of steady-state MSD of algorithms are recognized in Table 3. A time-invariant system is modeled, and its coefficients vector of the unknown system is $\mathbf{W}_O = [0.8, 0.2, -0.7, 0.2, 0.1]^T$. A time-varying system is modeled, and its coefficients are varied from a random walk process defined by $\mathbf{W}_O(n) = \mathbf{W}_O + \boldsymbol{\tau}(n)$, where $\boldsymbol{\tau}(n)$ is an i.i.d. Gaussian sequence with $E[\boldsymbol{\tau}(n)] = 0$ and $\sigma_\tau^2 = 0.01$. $\mathbf{W}_O = [0.8, 0.2, -0.7, 0.2, 0.1]^T$. The NLMF ($\mu$=0.005) [13], VSSLMFQ ($\alpha = 0.997$, $\gamma = 0.000002$, $a = 0.95$, $b = 0.995$, $\mu_{max} = 0.005$ and $\mu_{max} = 0$) [24], and OPLMF ($\gamma = 0.98$) algorithms

*For Time-invariant system*

The system noise is a Uniform distributed noise over the interval (0, 1) and the input signal is correlated input signal $y(n)$.

*Experiment 1*. *SNR*=3dB, The MSD curves for the NLMF [13], VSSLMFQ [24], and OPLMF algorithms with the correlated input signal are shown in Fig. 1(A). Fig. 1(B) shows the MSD error curves for the OPLMF algorithm. In order to make case of the OPLMF algorithm more persuasive, we provide a plot describing the evolution of $\mu(n)$ as a function of the number of iterations in Fig. 1(C).

*Experiment 2*. *SNR=1.5dB*. The MSD curves for the NLMF [13], VSSLMFQ [24], and OPLMF algorithms with the correlated input signal are shown in Fig. 2(A). Fig. 2(B) shows the MSD error curves for the OPLMF algorithm. In order to make case of the OPLMF algorithm more persuasive, we provide a plot describing the evolution of $\mu(n)$ as a function of the number of iterations in Fig. 2(C).

*Experiment 3*. *SNR=0dB*. The MSD curves for the NLMF [13], VSSLMFQ [24], and OPLMF algorithms with the correlated input signal are shown in Fig. 3(A). Fig. 3(B) shows the MSD error curves for the OPLMF algorithm. In order to make case of the OPLMF algorithm more persuasive, we provide a plot describing the evolution of $\mu(n)$ as a function of the number of iterations in Fig. 3(C).

*For time-varying system*

*Experiment 4*. *SNR=1dB, Gaussian noise, and uncorrelated input signal*
A time-varying system is modeled. The MSD curves for the NLMF [13], VSSLMFQ [24], and OPLMF algorithms with the uncorrelated input signal are shown in Fig. 4(A). Fig. 4(B) depicts the MSD error curves for the OPLMAT algorithm. In order to make case of the OPLMAT algorithm more persuasive, we provide a plot elucidating the evolution of $\mu(n)$ as a function of the number of iterations in Fig. 4(C).

*Experiment 5*. *SNR=1dB, Binary distribution noise, and correlated input signal*
A time-varying system is modeled. The MSD curves for the NLMF [13], VSSLMFQ [24], and OPLMF algorithms with the correlated input signal are shown in Fig. 5(A). Fig. 5(B) plots the MSD error curves for the OPLMF algorithm. In order to make case of the OPLMF algorithm more persuasive, we provide a plot showing the evolution of $\mu(n)$ as a function of the number of iterations in Fig. 5(C).

*Experiment 6*. *SNR=0dB, Rayleigh distribution noise, and correlated input signal*
Rayleigh distribution with 3 and the input signal is correlated input signal $y(n)$. A time-varying system is modeled. The MSD curves for the NLMF [13], VSSLMFQ [24], and OPLMF algorithms with the correlated input signal are shown in Fig. 6(A). Fig. 6(B) shows the MSD error curves for the OPLMF algorithm. In order to make case of the OPLMF algorithm more persuasive, we provide a plot showing the evolution of $\mu(n)$ as a function of the number of iterations in Fig. 6(C).

*Experiment 7*. *SNR=3dB, Poisson distribution noise, and correlated input signal*
Poisson distribution with 1 and the input signal is correlated input signal $y(n)$. A time-varying system is modeled. The MSD curves for the NLMF [13], VSSLMFQ [24], and OPLMF algorithms with the correlated input signal are shown in Fig. 7(A). Fig. 7(B) shows the MSD error curves for the OPLMF algorithm. In order to make case of the OPLMF algorithm more persuasive, we provide a plot showing the evolution of $\mu(n)$ as a function of the number of iterations in Fig. 7(C).

Fig.1(A), Fig. 2(A), Fig. 3(A), Fig. 4(A), Fig. 5(A), Fig. 6(A) and Fig. 7(A) show the OPLMF algorithm has a smaller misalignment than the NLMF and VSSLMFQ algorithms in steady-state stage. The reason for this observation is small $\mu(n)$ in steady-state stage. From Fig. 1(A) and Table 3, the improvement due to implementation of the NLMF and VSSLMFQ algorithms can nearly approach 27.65dB and 24.63dB,

respectively. Thus, compared to the NLMF and VSSLMFQ algorithms, the OPLMF algorithm can perform better in identifying the unknown coefficients under this condition. From the other experimental results, the same conclusion can be obtained.

Fig. 1(B), Fig. 2(B), Fig. 3(B), Fig. 4(B), Fig. 5(B), Fig. 6(B) and Fig. 7(B) show the MSD error. We observe an excellent match between predictions provided by our newly designed algorithm and results given by Monte Carlo simulations.

From Fig. 1(C), Fig. 2(C), Fig. 3(C), Fig. 4(C), Fig. 5(C), Fig. 6(C) and Fig. 7(C), we know that $\mu(n)$ has a large value in the initial stage, which results in a high convergence rate as expected. After the algorithm converges on the point where a low misalignment is desired, $\mu(n)$ decrease automatically.

From the results of *Experiment* 1- *Experiment* 3, we observe that the OPLMF algorithm is more robust to SNR than the NLMF and VSSLMFQ algorithms. In addition, from the results of *Experiment* 4- *Experiment* 7, we observe that compared with the NLMF and VSSLMFQ algorithms, OPLMF algorithm is more robust to SNR, multiple types of non-Gaussian noises and different types of input signals.

Table 3. MSD values

|  | MSD / dB | | |
| --- | --- | --- | --- |
|  | NLMF | VSSLMFQ | OPLMF |
| Experiment 1 | -28.51 | -31.53 | -60.91 |
| Experiment 2 | -26.17 | divergence | -57.24 |
| Experiment 3 | -30.82 | -33.84 | -58.47 |
| Experiment 4 | -16.88 | -21.06 | -44.37 |
| Experiment 5 | -22.43 | -26.41 | -57.28 |
| Experiment 6 | -18.50 | -26.92 | -85.88 |
| Experiment 7 | -19.25 | -22.69 | -28.34 |

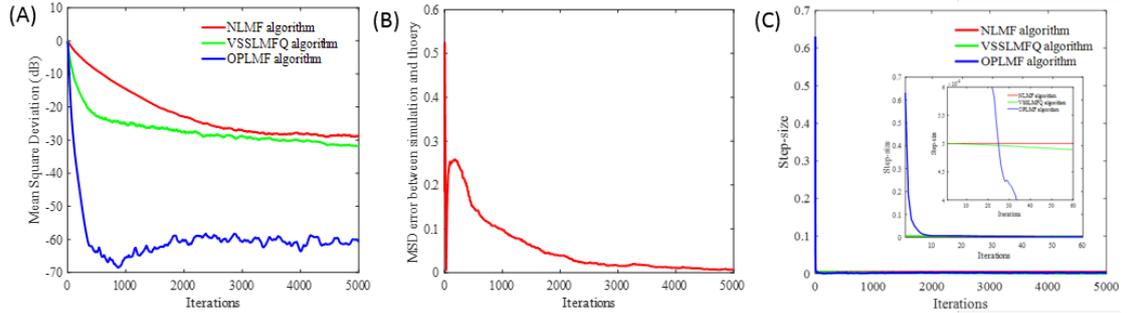

Fig.1. Comparison of MSD under in *Experiment 1*. (A) MSD comparisons result. These three curves represent the trend of the estimation error of NLMF (read), VSSLMFQ (green) and OPLMF (blue) with the number of iterations, respectively; (B) Learning curves of MSD error. This curve represent the trend of (MSD(*n*)-Theory (Eq.(30)) with the iteration; (C) Evolution of *μ(n)* as a function of iteration number. These three curves represent the trend of the step-size of NLMF (read), VSSLMFQ (green) and OPLMF (blue) with the number of iterations, respectively.

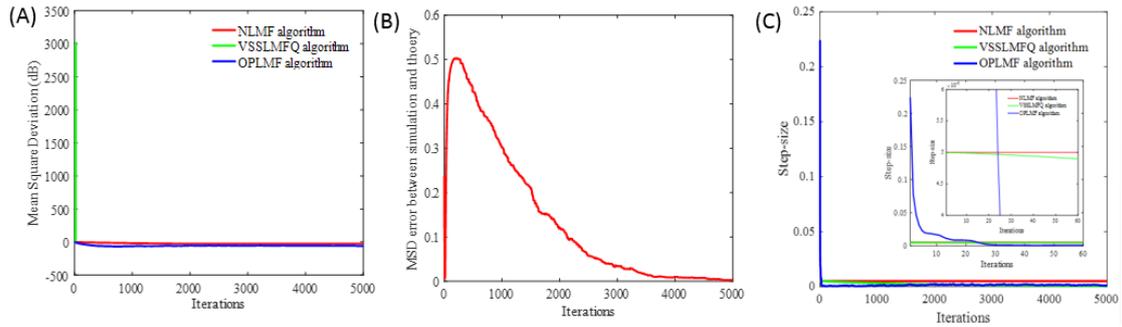

Fig.2. Comparison of MSD under in *Experiment 2*. (A) MSD comparisons result. These three curves represent the trend of the estimation error of NLMF (read), VSSLMFQ (green) and OPLMF (blue) with the number of iterations, respectively; (B) Learning curves of MSD error. This curve represent the trend of (MSD($n$)-Theory (Eq.(30)) with the iteration; (C) Evolution of $\mu(n)$ as a function of iteration number. These three curves represent the trend of the step-size of NLMF (read), VSSLMFQ (green) and OPLMF (blue) with the number of iterations, respectively.

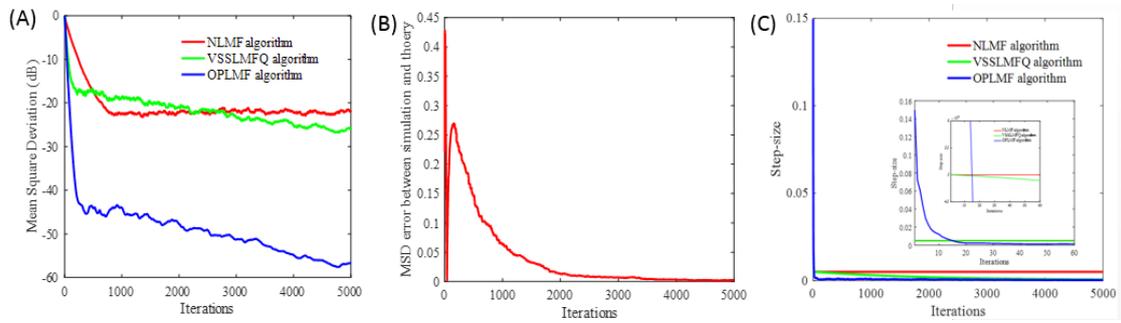

Fig.3. Comparison of MSD under in *Experiment 3*. (A) MSD comparisons result. These three curves represent the trend of the estimation error of NLMF (read), VSSLMFQ (green) and OPLMF (blue) with the number of iterations, respectively; (B) Learning curves of MSD error. This curve represent the trend of (MSD($n$)-Theory (Eq.(30)) with the iteration; (C) Evolution of $\mu(n)$ as a function of iteration number. These three curves represent the trend of the step-size of NLMF (read), VSSLMFQ (green) and OPLMF (blue) with the number of iterations, respectively.

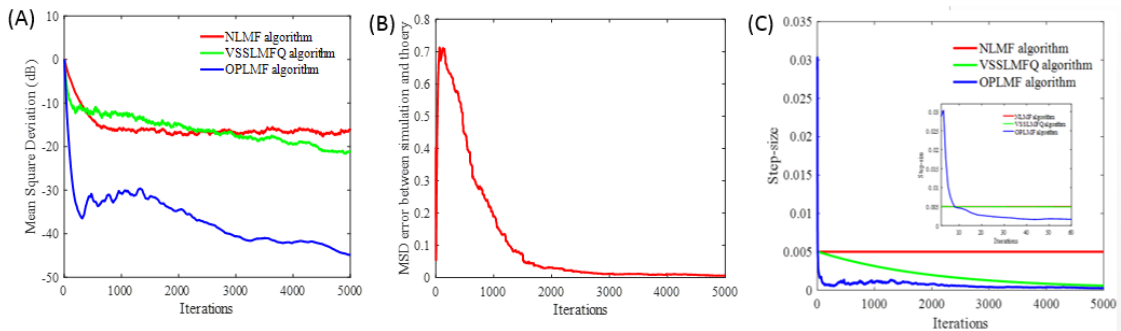

Fig.4. Comparison of MSD under in *Experiment 4*. (A) MSD comparisons result. These three curves represent the trend of the estimation error of NLMF (read), VSSLMFQ (green) and OPLMF (blue) with the number of iterations, respectively; (B) Learning curves of MSD error. This curve represent the trend of (MSD($n$)-Theory (Eq.(30)) with the iteration; (C) Evolution of $\mu(n)$ as a function of iteration number. These three curves represent the trend of the step-size of NLMF (read), VSSLMFQ (green) and OPLMF (blue) with the number of iterations, respectively.

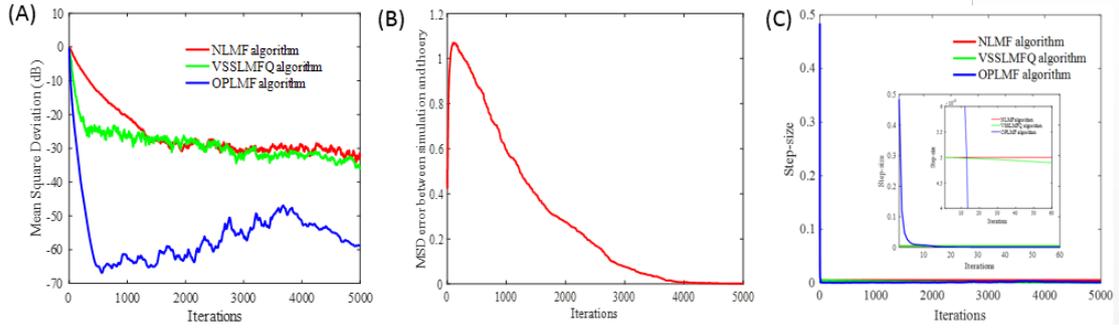

Fig.5. Comparison of MSD under in *Experiment 5*. (A) MSD comparisons result. These three curves represent the trend of the estimation error of NLMF (read), VSSLMFQ (green) and OPLMF (blue) with the number of iterations, respectively; (B) Learning curves of MSD error. This curve represent the trend of (MSD($n$)-Theory (Eq.(30)) with the iteration; (C) Evolution of $\mu(n)$ as a function of iteration number. These three curves represent the trend of the step-size of NLMF (read), VSSLMFQ (green) and OPLMF (blue) with the number of iterations, respectively.

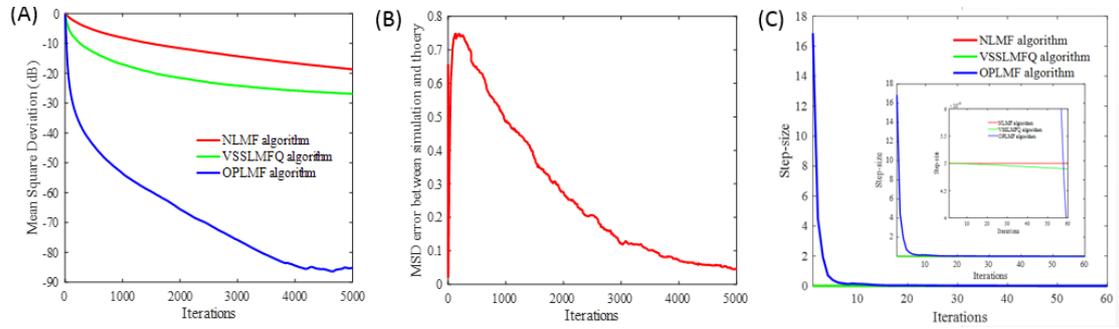

Fig.6. Comparison of MSD under in *Experiment 6*. (A) MSD comparisons result. These three curves represent the trend of the estimation error of NLMF (read), VSSLMFQ (green) and OPLMF (blue) with the number of iterations, respectively; (B) Learning curves of MSD error. This curve represent the trend of (MSD($n$)-Theory (Eq.(30)) with the iteration; (C) Evolution of $\mu(n)$ as a function of iteration number. These three curves represent the trend of the step-size of NLMF (read), VSSLMFQ (green) and OPLMF (blue) with the number of iterations, respectively.

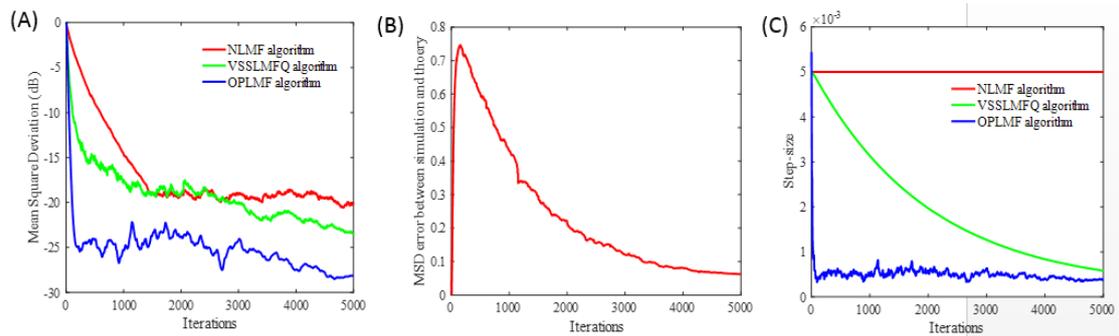

Fig.7. Comparison of MSD under in *Experiment 7*. (A) MSD comparisons result. These three curves represent the trend of the estimation error of NLMF (read), VSSLMFQ (green) and OPLMF (blue) with the number of iterations, respectively; (B) Learning curves of MSD error. This curve represent the trend of (MSD($n$)-Theory (Eq.(30)) with the iteration; (C) Evolution of $\mu(n)$ as a function of iteration number. These three curves represent the trend of the step-size of NLMF (read), VSSLMFQ (green) and OPLMF (blue) with the number of iterations, respectively.

## 5 Conclusions

In the context of system identification using the LMF-type algorithm, this paper described a novel way to derive an optimal variable step-size of the LMF algorithm (OPLMF) based on mean-square deviation (MSD) analysis. The OPLMF algorithm is developed in this paper in order to addresses both time-varying systems and systems where many types of noise (Gaussian, Uniform, Binary, Rayleigh and Poisson distributions noise) exists simultaneously under low SNR. Our method could achieve a very low steady-state error theoretically, with the step-size updated based on the number of iteration, and a faster convergence rate than prior methods. Simulation results showed that the proposed algorithm performed better than the NLMF and VSSLMFQ algorithms under different scenarios involving both time-invariant and time-varying systems, as well as under different noise distributions with low SNR. Both theoretical analysis and simulations provided corroborated results.

**Acknowledgments** This work was supported in part by the National Natural Science Foundation of China (NSFC) under Grants 61871420.

## References

[1] A.H. Sayed. Adaptive Filters [M]. Wiley, Hoboken, 2008.

[2] Guan S, Zhi L. Normalised spline adaptive filtering algorithm for nonlinear system identification [J]. Neural Processing Letters, 2017, 46(5):595-607.

[3] Zhi L, Guan S. Diffusion normalized Huber adaptive filtering algorithm [J]. Journal of the Franklin Institute, 2018, 355(8):3812-3825.

[4] Chen H, Chen J, Muir L A, et al. Functional organization of the human 4D Nucleome [J]. Proceedings of the National Academy of Sciences of the United States of America, 2015, 112(26):8002-7.

[5] Hitziger S, Clerc, Maureen, Saillet, Sandrine, et al. Adaptive waveform learning: a framework for modeling variability in neurophysiological signals [J]. IEEE Transactions on Signal Processing, 2017, 65(16):4324-4338.

[6] Walach E, Widrow B. The least mean fouth (LMF) adaptive algorithm and its family [J]. IEEE Transactions on Information Theory, 1984, 30(2):275-283.

[7] Sihai GUAN, Zhi Li. Nonparametric variable step-size LMAT algorithm [J]. Circuits Systems & Signal Processing, 2017, 36(3):1322-1339.

[8] Ciochină S, Paleologu C, Benesty J. An optimized NLMS algorithm for system identification [J]. Signal Processing. 2016, 118:115-121.

[9] Guan S, Li Z. Optimal step size of least mean absolute third algorithm [J]. Signal, Image and Video Processing, 2017, 11(6):1105-1113.

[10] Hubscher P I, Bermudez J C M. An improved statistical analysis of the least mean fourth (LMF) adaptive algorithm [J]. IEEE Transactions on Signal Processing, 2003, 51(3):664-671.

[11] Nascimento V H, Bermudez J C M. Probability of divergence for the least-mean

[11] fourth algorithm [J]. IEEE Transactions on Signal Processing, 2006, 54(4):1376-1385.

[12] Hubscher P I, Bermudez J C M, et al. A mean-square stability analysis of the least mean fourth adaptive algorithm [J]. IEEE Transactions on Signal Processing, 2007, 55(8):4018-4028.

[13] Eweda, Bershad. Stochastic analysis of a stable normalized least mean fourth algorithm for adaptive noise canceling with a white Gaussian reference [J]. IEEE Transactions on Signal Processing, 2012, 60(12):6235-6244.

[14] Eweda E. Dependence of the stability of the least mean fourth algorithm on target weights non-stationarity [J]. IEEE Transactions on Signal Processing, 2014, 62(7):1634-1643.

[15] Navia-Vazquez A, Arenas-Garcia J. Combination of recursive least p-norm algorithms for robust adaptive filtering in alpha-stable noise [J]. IEEE Transactions on Signal Processing, 2012, 60(3):1478-1482.

[16] Aydin, G., Arikan, O., Cetin, A.E. Robust adaptive filtering algorithms for α-stable random processes [J]. IEEE Trans. Circuits Syst. II Analog Digit. Signal Process. 1999, 46(2):198-202.

[17] Zerguine A. Convergence and steady-state analysis of the normalized least mean fourth algorithm [J]. Digital Signal Processing, 2007, 17(1):17-31.

[18] Zerguine A, Chan M K, et al. Convergence and tracking analysis of a variable normalised LMF (XE-NLMF) algorithm [J]. Signal Processing, 2009, 89(5):778-790.

[19] Eweda E. Global stabilization of the least mean fourth algorithm [J]. IEEE Transactions on Signal Processing, 2012, 60(3):1473-1477.

[20] Shin H C, Sayed A H, et al. Variable step-size NLMS and affine projection algorithms [J]. IEEE Signal Processing Letters, 2004, 11(2):132-135.

[21] Zerguine A, Moinuddin M, et al. A noise constrained least mean fourth (NCLMF) adaptive algorithm [J]. Signal Processing, 2011, 91(1):136-149.

[22] Bershad N J, José C.M. Bermudez. Mean-square stability of the normalized least-mean fourth algorithm for white Gaussian inputs [J]. Digital Signal Processing, 2011, 21(6):694-700.

[23] Zhao S, Man Z, Khoo S, et al. Variable step-size LMS algorithm with a quotient form [J]. Signal Processing, 2009, 89(1):67-76.

[24] Syed Muhammad Asad, et al. A robust and stable variable step-size design for the LMF using quotient form [J]. Signal Processing, 2019, 162:196-210.

[25] Benesty J, Rey H, Vega L R, et al. A nonparametric VSSNLMS algorithm [J]. IEEE Signal Processing Letters, 2006, 13(10):581-584.

[26] Spiegel, M.R. Mathematical handbook of formulas and tables [M]. McGraw-Hill, New York, 2012.